\documentclass[10pt,oneside]{article}
\usepackage[english]{babel}
\usepackage{geometry} 
\usepackage{graphicx}
\usepackage{graphics}
\usepackage{amsfonts}
\usepackage{amsmath}
\usepackage{hyperref}

\title{Statistical Agent Based Modelization\\ of the Phenomenon of Drug Abuse}

\author{Riccardo Di Clemente$^{1,2}$\thanks{\mbox{Corresponding author. \emph{E-mail~address}:~riccardo.diclemente@imtlucca.it}} \& Luciano Pietronero$^{1,3}$ \\
\footnotesize
$^{1}$\textit{Istituto dei Sistemi Complessi-CNR, Via dei Taurini 19, 00185 Rome, Italy}\\
\footnotesize
$^{2}$\textit{IMT Institute for Advanced Studies Lucca, Piazza S. Ponziano, 6 55100, Lucca, Italy}\\
\footnotesize
$^{3}$\textit{Universit\`a``La Sapienza'', Dip. di Fisica, Piazz.le Aldo Moro 2, 00185, Rome, Italy}}

\begin{document}
\date{Published 25-July-2012 on Scientific Reports 2, 532  \href{http://dx.doi.org/10.1038/srep00532}{DOI: 10.1038/srep00532} (2012)}
\maketitle

\begin{abstract}
We introduce a statistical agent based model to describe the phenomenon of drug abuse and its dynamical evolution at the individual and global level.
The agents are heterogeneous with respect to their intrinsic inclination to drugs, to their budget attitude and social environment.
 The various levels of drug use were inspired by the professional description of the phenomenon and this permits a direct comparison with all available data.
We show that certain elements have a great importance to start the use of drugs, for example the rare events in the personal experiences which permit to overcame the barrier of drug use occasionally.
The analysis of how the system reacts to perturbations is very important to understand its key elements  and it provides strategies for effective policy making.
The present model represents the first step of a realistic description of this phenomenon and can be easily generalized in various directions.

\end{abstract}

\section*{Introduction}
The ambition to develop a quantitative description of people's behavior and introduce novel ideas and methods in socio-economic disciplines \cite{1,1a,1b} (see also the future ict project www.futurict.eu) is one of the main challenges of statistical physics and complexity theory. Agent based models (ABMs) \cite{2,3} represent a broad framework to address these questions. 
They can describe some of the most important properties that get inspiration from physical phenomena. 
These are for example the importance of heterogeneity, large (critical) fluctuations, self-organized criticality and the lack of cause-effect relation. 
The the key features of ABM is the suitable choice of the nature of the agents, their interactions and their dynamical evolution.

In this paper, we present an Agent Based Model which aims to describe, at a reasonably microscopic level, the phenomenon of drug abuse, its evolution and control.
The scheme of the model was inspired by the participation of one of the authors (L.P.) to the national initiative PREVO.LAB \cite{4,4a,4c} whose objective is to analyze and control the phenomenon of drug abuse in Italy.
Therefore, the model is closely aligned with the professional analysis in this field and permits a direct comparison of concepts and parameters with those actually observed and analyzed in reality. 
In this respect, the model is rather realistic and suitable for direct applications, including control, forecast of the phenomenon and for policymaking.
The information in this field is very scattered and ranges from highly accurate information about people who become hospitalized to the little known detail about those at the beginning of the process. 
The model aim is to provide a complete framework to describe the phenomenon, whose parameters are fixed by the best known facts.
 In this way, the model is able to extrapolate the knowledge of the less known submerged elements.
This Agent Based Model, with its parameters fixed by real observations, exhibits the following features: 

\begin{itemize}
\item The tail fluctuations of personal experience (environment and interactions) are critical components for overcoming the natural barrier to start drug consumption, for the first time. This also implies that the heterogeneous nature of the social-network is very important and cannot be represented by an average situation.
\item The economic barrier plays a relatively minor role with an appreciable importance only at the beginning.
\item The model permits to track the individual history of an agent and in turn, this information could be directly compared with medical and other data.
\item One can analyze the response of the system to external changes of some parameters. This can help optimizing the control and defining a suitable policy.
\item The model is flexible and can be easily improved by introducing specific elements that may be inspired by new observations. For example, it could be implemented in a specific social complex network \cite{17} and one can also introduce different drugs with different market socioeconomic characteristics, related for example to the age, the economic background or the gender.
\end{itemize}

In summary, we introduce an ABM  with variables and parameters defined in a way that makes efficient use of all data and information available in the field. This permits us to achieve a complete description of the phenomenon both at the individual and global level, and to identify its crucial elements and its responsiveness to changes in any of its parameters.
 
The study of drug consumption and trend predictions began in the '70s with the analysis of trend historical data of New York City \cite{5}. In the '90s Everingham and Rydell \cite{6} proposed a Markovian process to analyze and predict cocaine consumption in the USA. The first ABM model introduced was \textit{Drugmart} by Agar and Wilson \cite{7,8}, revised by the UK Office of Science and Technology report \cite{9}, which is a  simple network model. Another more complex model is the Australian \textit{SimDrug} \cite{10} which simulates the consumption of heroin in Melbourne with accurate people's dynamic reconstruction. 

With respect to these preliminary models, our ABM has a much higher level of complexity and realism and it provides information on all elements  of the phenomenon both at the individual and global level.
To this end,  it is important to note that our main conclusions could not have been derived from these previous models.

\section*{Results}

\subsection*{The Model}

In our model, an agent $i$, with $i\in[1;N]$, is characterized by a number of parameters and interacts with the environment and with the other agents at each time $t\in[1;T]$. In the dynamic the unit time step will correspond to one week.
We now briefly introduce the key elements of the model, which are described in more detail in the methods section.

\textbf{Age} $y_{i}(t)=[10:100]$ years. The starting distribution of the agents' age  at $t_{0}$ is taken from the Italian data of ISTAT. In a first stage, we consider the property of a static situation  $y_{i}(t)=y_{i}(t_{0})$. Later, we also introduce the dynamic evolution of the system by increasing the agents' age with time.

\textbf{Consumers type} $S_i(t)=0;1;2;3;4$ defines the consumption stage and the level of dependence reached by a consumer: non users $(S_{i}=0)$; mini user with age $y_i(t)<26$ $(S_{i}=1)$; occasional user $(S_{i}=2)$; frequent user not pathological $(S_{i}=3)$; heavy user pathological $(S_{i}=4)$.
These categories are inspired by the professional classification of drug abuse from ``Diagnostic and Statistical Manual of Mental Disorders''  \cite{11}.
The mini user with $y_{i}<26$ are agents with low income that can be in contact with the normal dose and with a lower quantity of drug, the mini dose. This mini dose, as we will see later, was introduced by drug cartel with the purpose of stimulating in the young consumers a possible future dependency on drug \cite{4,4a}. 

\textbf{Personal budget and saving behavior}.
 Behavioral analysis \cite{4a} has shown that the money a person would spend on initial drug consumption is rather related to personal saving attitude for leisure opportunities, than to the total personal budget. 
Therefore, the only variation we make in the budget $m_{i}(y_{i}(t))$ refers to the difference between the adult working population with age $y_i(t)\ge26$ with $m_{i}(y_{i}(t))=1$, and  young unemployed or student population with age $y_i(t)<26$ with  $m_{i}(y_{i}(t))=0.2$. The threshold of age $26$ represents the average age of the change of spending and work habits \cite{4a}.
 The crucial point will be the parameter  $\gamma_i\in[0;\infty]$ (saving behavior), which describes the tendency of an agent to save or spend money. Inspired by various indicators of social behavior, we assign to it a  lognormal distribution \cite{12} with variance  $\Gamma_\sigma$ and average $\Gamma_\mu$.  A value of $\gamma_i\approx0$ means that the agent is inclined to spend his leisure budget while $\gamma_i\ge1$ corresponds to a strong willingness to save money and do not use it for buying drugs. 
 
\textbf{Personal exposure}  $e_i$ describes every possible contact between the agent and the drug world, in particular the input the agent receives about drug consumption. In the present model we adopt a simplified model of interaction among agents but the generalization to a complex social network would be a natural extension in the model \cite{17,15,16}.
  
\textbf{Inclination to drug}   $\beta_i$  controls the intrinsic (genetic, educational, cultural,  etc.) agent's barrier towards drugs. Also in this case, inspired by the social studies \cite{13}, we adopt lognormal distribution with  $\sigma=B_\sigma,\mu=B_\mu$.  
For every agent we define the barrier to be overcome in order to start using drugs: $b_i^*(t)=\beta_i/\alpha(y_i(t))$; where $\alpha(y_i(t))$ is a function dependent on the agent's age and it has the characteristic to increase the barrier for older agents.
This function is described by a simple analytical form which reproduces the data of Prevo.Lab \cite{4}.
 
\textbf{Drug addiction}: describes how an agent becomes addicted to drugs and how this modifies her behavior. This property is defined by the function $d_i(\pi_i(t),S_i(t))$ which can increase or decrease $b_i^*(t)$, depending on the persistence of the agent in a given stage, where $\pi_{i}(t)$ is an exponential function of consecutive drug assumption.
The inclusion of this effect redefines the drug dependence function: $b_{i}(t)=b_{i}^{*}(t)-d_{i}(\pi_{i}(t),S_{i}(t))$.

In the drug market there are two different dose levels: mini-dose and dose. The first has a lower quantity and a cheaper price. It is used for attracting young people to drug consumption while the second represents the normal dose for adults and for young usual users. 
The mini-dose is very important for the analysis of drug consumption because it is often the way in which the young generations start the drug use. Moreover it represents the cheapest and easiest way of transmission of drug among young people \cite{4a}.
In our model the mini-dose could symbolize a light drug that can lead to the intake of the normal one.
The mini-dose price $p_m(t)$ is supposed to be substantially smaller than a dose. We have chosen an lower magnitude order as reasonable estimation of the mini dose price which is about one tenth of the dose \cite{4}. The price of the dose $p_d(t)$ is based on a simplified relation between supply and demand. In our case we assume that the total supply is fixed so the price is simply due to the total number of users.

\begin{equation}
\label{pricedose}
\begin{array}{ccc}
p_d(t)=\frac {\sum\limits^{N}_{k=1}\delta{\scriptstyle (S_{k}(t)=1,2,3,4)}}{N}&\quad\quad&p_m(t)=\frac{p_d(t)}{10}
\end{array}
\end{equation}
This relation could be easily made more realistic with a variable level of total drug amount.  

The agents interact and evolve through the comparison of the two major barriers: the \emph{Economic Barrier} ``EB". and the \emph{Socio-Emotional Barrier} ``SEB".
The first barrier compares the price of the drug with the buying capacity and it is defined by:
\begin{equation}
\label{eq: acquisto droga}
\left\{ 
\begin{array}{lcc}
\frac{p_{m}(t)}{m_{i}(y_{i}(t))}<1-\gamma_{i}\;&\mbox{ if }&\; y_{i}(t)<26\\
\frac{p_{d}(t)}{m_{i}(y_{i}(t))}<1-\gamma_{i}\;&\mbox{ if }&\; y_{i}(t)\ge26
\end{array}\right.
\end{equation}

The second barrier has the role of describing the social opportunities and the emotional process involved in drug consumption.
The social opportunities are defined in terms of the events by which the agents are subjected during a time unit.
 We describe this \textit{``daily noise''} by $r_i(t)$, random number drawn each time for each agent from a Gaussian distribution, $R_\mu=0$ , $R_\sigma$. Our choice to use the Gaussian distribution is due in large part to the lack of quantitative information that is available. It was also chosen because of the symmetric shape that could lead to a daily event that could generate a positive or negative effect in the drug intake. When more information becames available, it will be easy to modify accordingly.
 The condition to overcame the SEB is defined by: $b_{i}(t)+r_{i}(t)<e_{i}(t)$.
 
We have now all the elements that define the dynamics and the interaction between the agents. In summary the most important parameters of the model are $\gamma_i$, $r_i(t)$ and $\beta_i$ which are respectively money saving behavior, daily noise and intrinsic inclination to drug. 

We can define the transition probability of each agent  from the stage  $S_i(t)$ to $S_i(t+1)$ as indicated in Figure 1a. The corresponding rates are described in detail in the methods section.

\subsection*{Tuning the parameters}

The first point we consider is to tune the model parameters to a realistic description of the observed data. A crucial parameter in this respect is the observed fraction of people using drugs.
In order to address this question we have made several simulations with $N=1000$ and $T=1000$ (the time unit $t$ is one week). The strategy is to fix two distributions and examine the change of drug using population during the variation of the third one. Then we repeat the operation for all three distributions. 
In Figure 1b.c.d we see the change depending on the three distributions. We choose to have in our population about $10\%$ of drug users (a realistic value considering all different drugs \cite{4,13a,14}, which could be adjusted when better data is available). 
When studying the real drug consumption, in addition to these total values,  it is important  to split the agents in four age ranges \cite{4} labeled as: young $(15-25)$, young-adult $(26-35)$, adult $(36-45)$ and senior $(45-100)$. Also, the individual percentage of users within the four age ranges is used to fix our parameters. These procedures define the three distributions, which we name  \textit{``Usual Distributions"}, UD:

\begin{equation}
\label{valorisoliti}
\left\{ 
\begin{array}{lcl}
\beta_{i}&\dashrightarrow &\mbox{lognormal}(B_\mu=1.2,B_{\sigma}=2.7)\\
\gamma_{i}&\dashrightarrow &\mbox{lognormal}(\Gamma_{\mu}=0.6,\Gamma_{\sigma}=0.9)\\
r_{i}(t)&\dashrightarrow &\mbox{gaussian}(R_\mu=0,R_{\sigma}=0.6)
\end{array}\right.
\end{equation}

Therefore, these optimized distributions provide a complete and realistic parametrization of our model which can now be used to investigate a variety of problems. We start by considering the individual personal history per agent.

\begin{figure}[ht]
\centering
\includegraphics[width=1\textwidth]{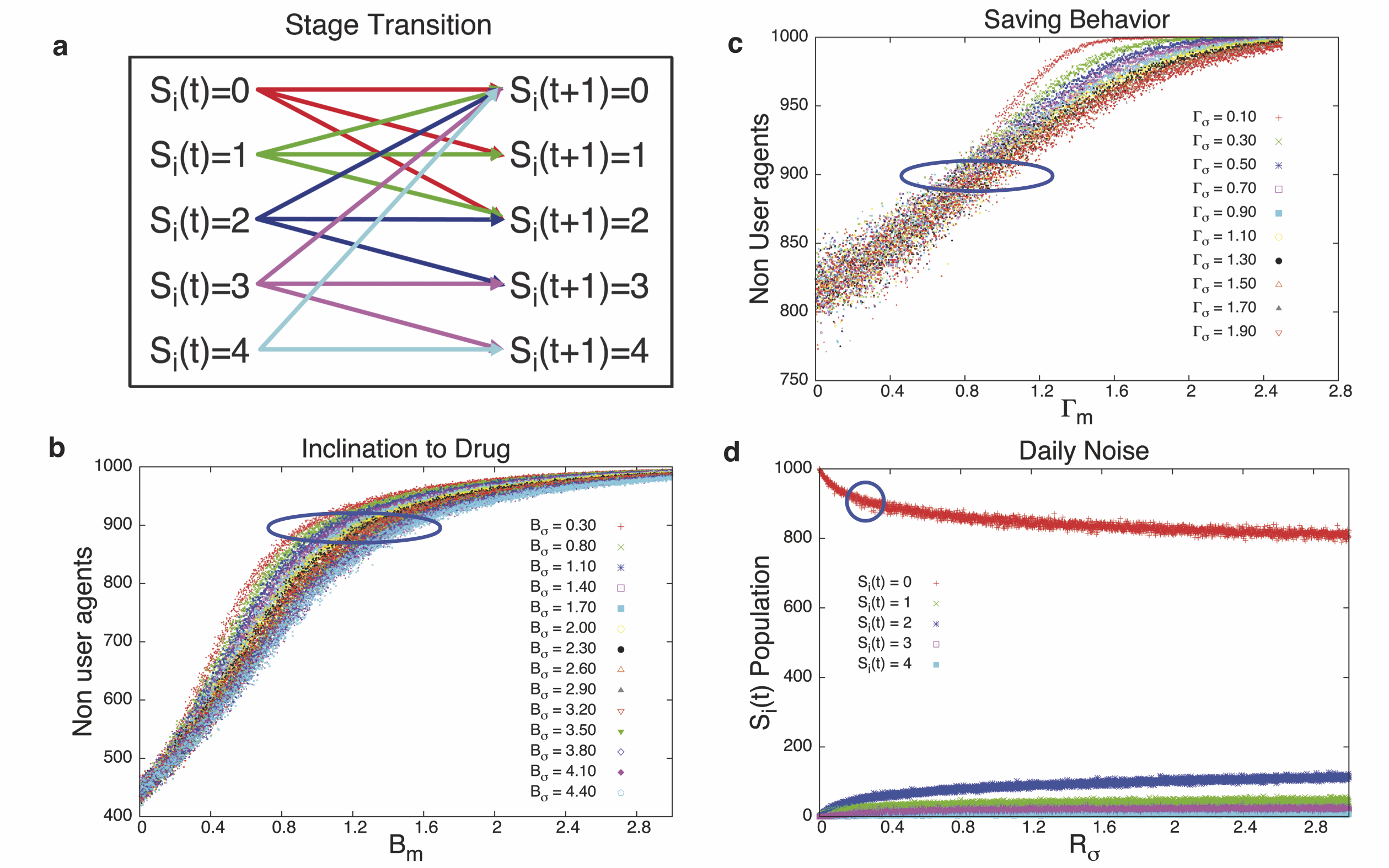}
\caption{\scriptsize\textbf{Stage transitions and parameter settings. a.} All possible stage transitions for an agent at stage $S_{i}(t)$. \textbf{b. c. d.} Simulations intended to select the main parameter of the model in relation to the realistic stage of the phenomenon (see text for more details). In particular in \textbf{b. } we show the effect of the parameter  $\beta_{i}$ (inclination to drug) distribution $(B_{m};B_{\sigma})$, in \textbf{c. } effect of a parameter $\gamma_{i}$ (saving behavior) distribution $(\Gamma_{m};\Gamma_{\sigma})$ and in \textbf{d.}  $r_{i}$ (daily noise) distribution $(R_{m}=0;R_{\sigma})$. The blue circles identify the realistic areas to fix the values of the three distributions so that they correspond to a total of $10\%$ of users. This parameter optimization refers also to the age ranges discussed in the text.}
 \end{figure}

\subsection*{Individual agent history}
It is interesting to note that the individual history can be vastly different in reality and we are going to observe that our ABM is able to reproduce a dynamic variety of agent's history. The table in Fig. 2a shows $\Delta t=68$ consecutive iterations of eight users, and it outlines some of the different users typology that can be produced by the model:

\begin{itemize}
\item $\#64$ is a senior agent that never tested drugs;
\item  $\#141$ is a young agent that, as an occasional consumer, tries the normal dose;
\item $\#1$(young-adult) and $\#343$(young) are two agents that are occasional consumers;
\item $\#15$(young-adult) and  $\#130$(young) are both heavy consumers that evolve into addicted;
\item $\#45$(adult) and $\#494$(young)  are two agents that try drugs on extremely rare occasions.
\end{itemize}

The fact that our model reproduces a realistic heterogeneity of agents' history is a very important element.
 This realistic variability is in fact crucial to have an accurate description of the phenomenon that could be compared to a considerable amount of real data. A possible upgrade would be analyzing and contrasting the real drug user clinical histories. By means of the above we could provide a probabilistic prediction of detoxification options for each clinical case.
In order to gain more insight in the individual evolution, we can observe the trend of the barrier for inclination to drugs use, the Social Emotional Barrier (SEB)  and the stage reached by an individual user $i$ at anytime $t$. 
The Figure 2b shows the history of a young user  $y_i=19$ with a low intrinsic inclination barrier $\beta_i$, for $\Delta t=300$ consecutive iterations. The user stage $S_i(t)$ is indicated in red dots, the left term of SEB inequality barrier in green, and in blue the starting value of inclination to drug $b^{*}_{i}(t)$. We can see that this agent develops all the stages with important fluctuations and finally ends up being an addict. At this level the green barrier line declines  strongly and finally we can also observe that the agent detoxifies. In the case above the effect of aging is not yet included $b^{*}_{i}(t)=b^{*}_{i}(t_{0})$. 
 
Now we consider a complete dynamic evolution in which aging is explicitly included. We set the year length $\tau=50t$, in which one iteration corresponds to a real week time spend. We elaborate a set of rules to define the death of an agent due to the aging and the birth of the new one in order to have a constant number of agents $N$. We also define a simple process, from the data \cite{4c,10,13a}, which describes the possibility that an agent can die of overdose and be replaced by a new one.
In Fig. 2c we reproduce the personal history of an agent in the aging process. The figure shows the increase of the barrier to $b^*_i(t)$ due to the agent's aging. The agent starts  with a high value of $\beta_i$ (low tendency to use drug) but the daily noise induces to try several drugs  $(S_i(t)=2)$. 
However, when the agent grows up, $b^*_i(t)$ rises, so the daily noise becomes less effective and finally the agent no longer uses drugs.
This example shows that the possibility to monitor a single agent's history can be an important tool to compare with real medical data.
Once again the future analysis could study the probability of stage change and compare it with real data from detoxification centers.

\begin{figure}[ht]
\centering
 \includegraphics[width=1\textwidth]{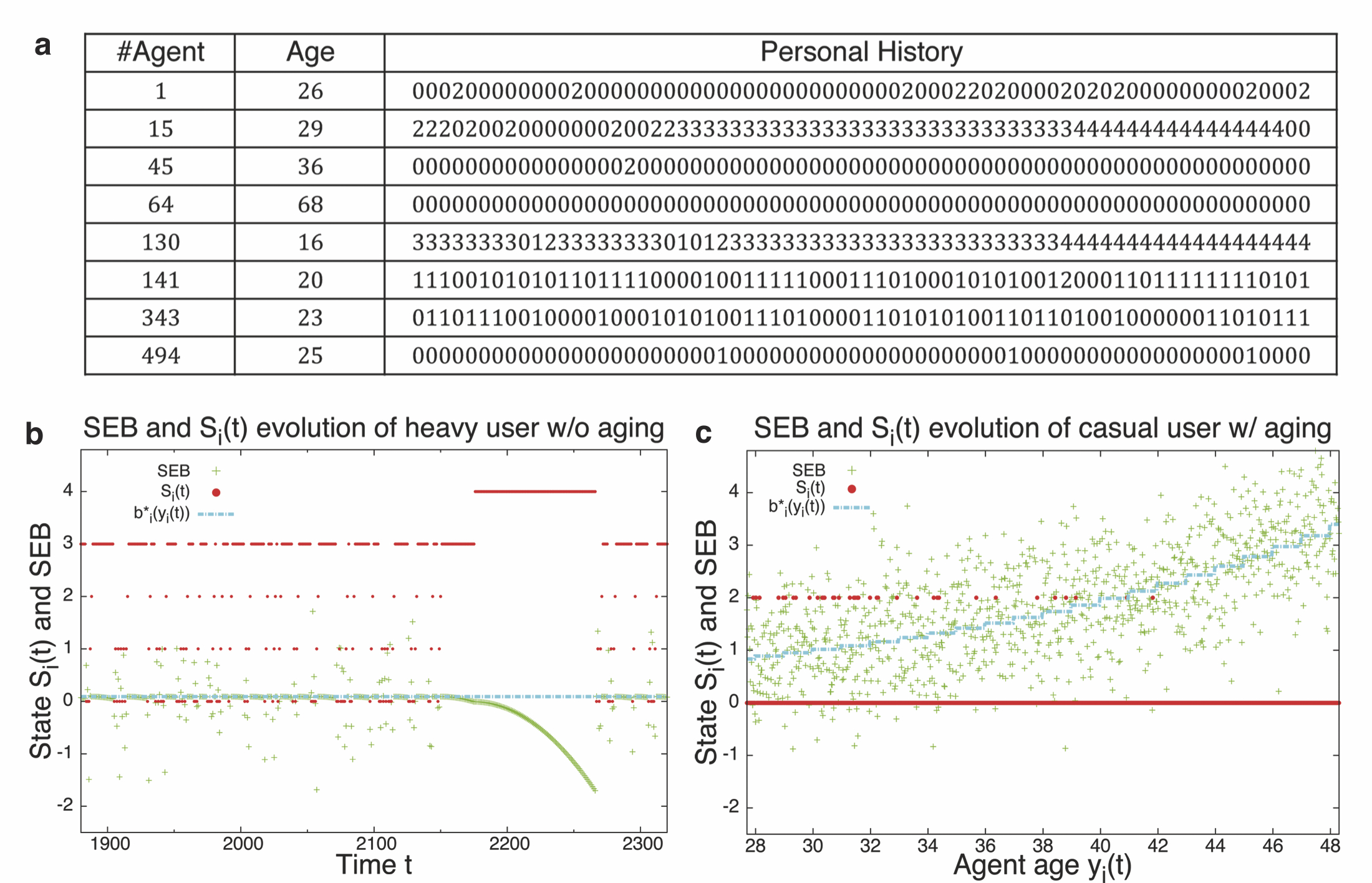}
 \caption{\scriptsize\textbf{Individual agent histories. a.} The table shows the histories of eight agents for $\Delta t=68$ consecutive times (One time unit corresponds to one week). One can see that our model is able to reproduce a variety of a realistic heterogeneity in agents' history. This history ranges: from the agent  $\#64$, who never tried drugs,  to agents characterized as heavy users $\#15;\#130$. \textbf{b.} Personal agent history without aging: the agent becomes a heavy user because after various fluctuations she reaches the stage $S_{i}(t)=4$. The green line describes the left term of Social Emotional Barrier and one can see that this barrier is drastically lower when the stage of dependence is reached. \textbf{c.} Personal agent history with aging. In this case the agent is a casual user, meaning that agent has frequently used drugs at a younger age. Then the barrier of inclination to drug use (blue line) $b^{*}_{i}(y_{i}(t))$ increases with the age. We can see that, in this way, the use of drugs $S_{i}(t)=2$ becomes more and more rare. Finally when the agent gets older the use of drug is completely eliminated. This example shows that the model provides a detailed description of individual histories, which could be directly compared to real data.
}
\end{figure}

\subsection*{Global property and effects of perturbations}

Our model also allows us to examine how the users population reacts to various sources stress, ranging from a different opportunity to be involved in drug use $(r_{i})$, a new agent's intrinsic education $(\beta_{i})$ or a change in saving behavior $(\gamma_{i})$.
 With the purpose of analyzing the agent's dynamic under these social modifications, we fixed two of the UD and analyzed the effect of changing  \textit{in-time}  of the third one. We will activate the \textit{in-time} modification between the time steps,  $t^{+}<t<2t^{+}$.
 
In Figure 3a we change the daily noise distribution from the UD to $R_{\mu}=0.3,R_{\sigma}=1.6$, with $t^{+}=4\tau$. The Figure shows how an increase of noise leads to an increase in the number of agents in the drug stage. As it is shown from the insert of Figure 3a the $r_{i}(t)$ distribution has different impact for each age range, given the dependency of   $b^*_{i}(t)$ on  age.

Figure 3b exhibits the change of the population stages $S_{i}(t)$ due to the modification of the intrinsic inclination toward drug use distribution $(B_{\mu}=1.2,B_{\sigma}=2.7)$ for the new born agent between $20\tau<t<40\tau$.  The agents who are born during the perturbation periods are more inclined to use drugs than the others. The insert of Figure 3b shows the trends between the age ranges and how the aging of the new agents increases temporally the number of drug users.

In Figure 3c we change the tendency of agent spending behavior. We replace the distribution of $\gamma_{i}$ from the UD with $(\Gamma_{m}=0.1;\Gamma_{\sigma}=0.3)$ for the new born between $20\tau<t<40\tau$. During this perturbation period the new born agents are more inclined to spend money for leisure, in this case drug. One can see from the insert of the picture the different trend among age ranges. According to the analysis, the saving behavior does not appear to have an important role in explaining drug assumption. The decrease in-time the $\gamma_{i}$ distribution produces a limited impact in the number of users compared to the effect $\beta_{i}$.

Finally  we focus on the role that mini users have on the introduction of drug among the new generations. This is an important point because it is hard to monitor and as an appreciable effect only after a long time. In Figure 4, we compare two situations of a population evolution. For the first $15\tau$ years the simulation does not consider the mini-dose. Consecutively  we included the mini user. We can see that the existence of mini user stage leads after many years to a large increase in the total number of drug users.

\begin{figure}[]
\centering
 \includegraphics[width=0.50\textwidth]{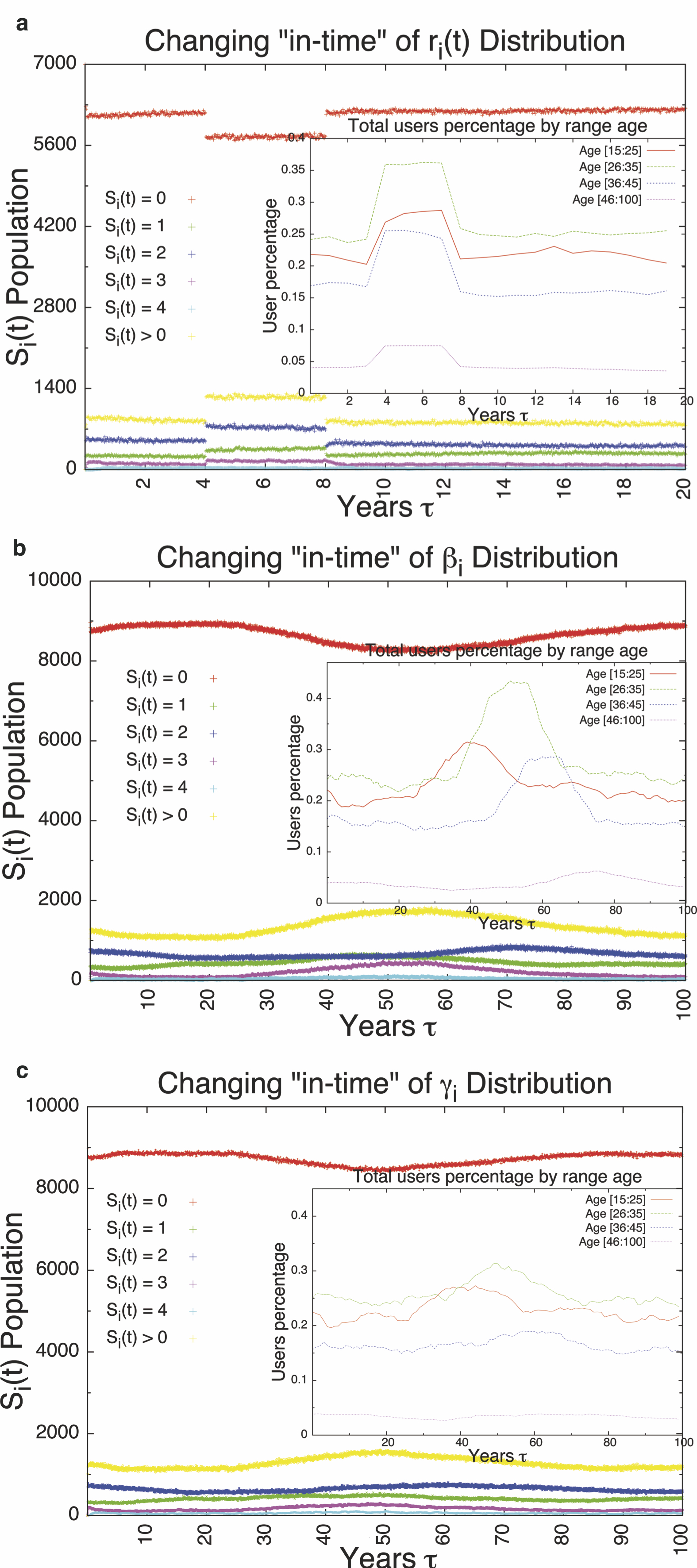}
 \caption{\scriptsize\textbf{Response of the system to external changes. a. } Fluctuation of the life style. Here we test how the system reacts to increased fluctuations for the rare events (tail of the $r_{i}(t)$ distribution)  that induce the starting of abuse. During the period $4\tau<t<8\tau$ the distribution of daily noise increased $(R_{m}=0.1;R_{\sigma}=1.3)$, the various ranges of population are affected by this change. In the insert we show how the total number of drug users is distributed among the various age ranges. From these studies, one can conclude that the system is strongly sensitive to this  tail fluctuations, which represents a very important point for the control of this phenomenon. 
\textbf{ b.} Cultural fluctuations. Here we show the effect of the changes in the genetic cultural barrier of inclination to drugs. We change the distribution of inclination to drug $\beta_{i}$ for the new agent during the period $20\tau<t<40\tau$ with $(B_{m}=1.2;B_{\sigma}=2.7)$. The lower intrinsic barrier leads to a creation of a new users generation more inclined to drugs abuse. We can see that the effect is not instantaneous and we can follow how it develops over the years. Also in this case, as we show in the insert,  one can see the large difference of age effect between the various age ranges. The pattern revealed can be directly compared to real data and is important for monitoring and controlling the phenomenon. 
\textbf{ c.} Fluctuations of saving money behavior. Here we examine the effect of the change of the saving behavior. We change the distribution of $\gamma_{i}$ for the new agent during the period $20\tau<t<40\tau$ with $(\Gamma_{m}=0.1;\Gamma_{\sigma}=0.3)$. A lower value of $\gamma_{i}$ means that an agent tends to spend money easily. In particular a value near to zero for an agent means that she has no problems for drug purchase. We can see that the effect is similar to the cultural fluctuation but it is of minor impact. The effect of spending capacity decreases with the age and in particular it does not affect the senior agents.}
 \end{figure}

\begin{figure}[ht]
\centering
 \includegraphics[width=0.70\textwidth]{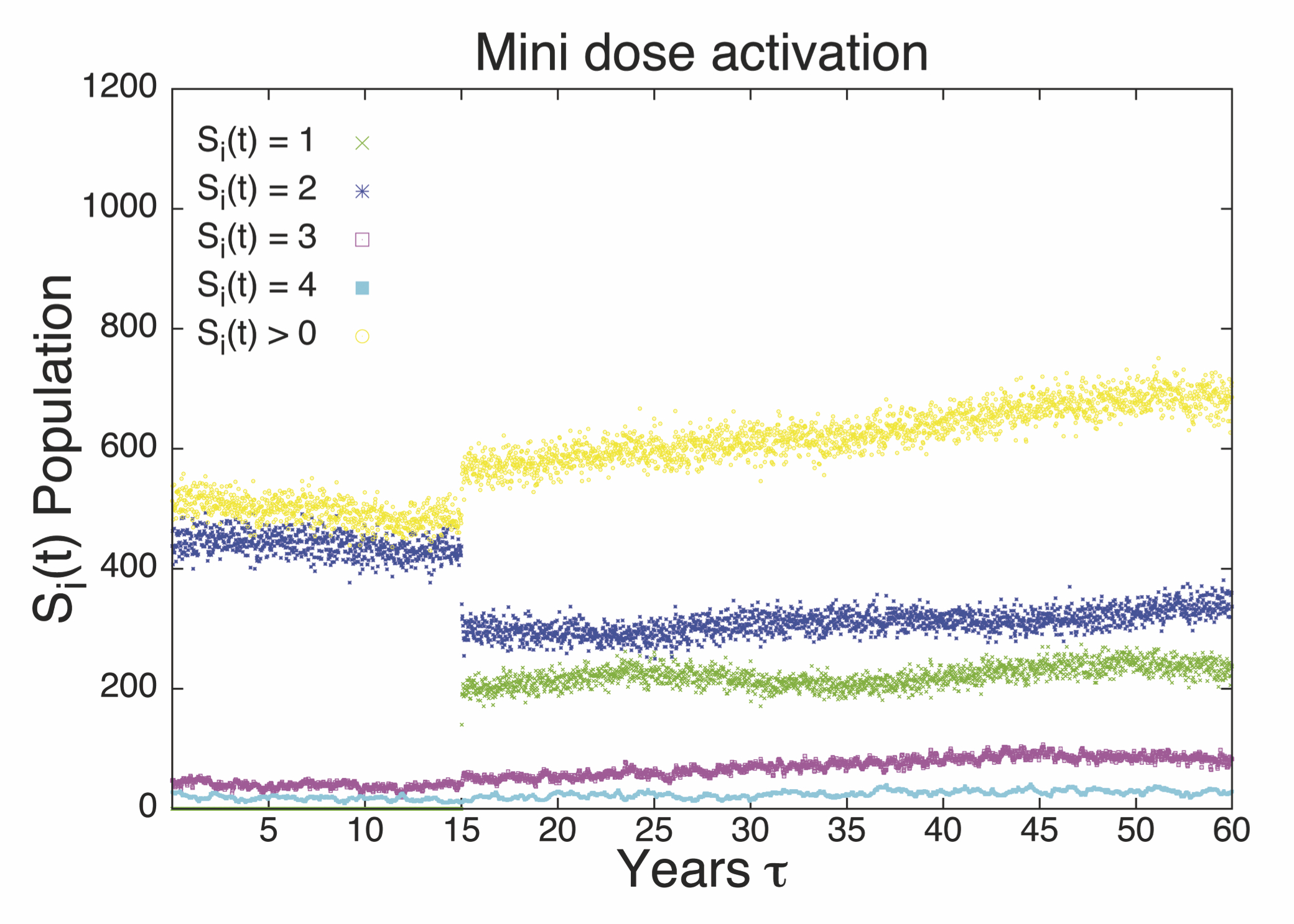}
 \caption{\scriptsize\textbf{Mini dose activation.}  We consider a population of $N=7000$ agents. At the beginning the market is without the mini dose, at a time $t=15\tau$ the mini dose is introduced in the system for young people. We only show the drug user stages and one can see an immediate change of the global users followed by a further increase of addict population. The results show that the effect of mini dose for young generation represents and extremely important element for the dynamics of the system.}
 \end{figure}

\section*{Discussion}

In conclusion, we have developed an agent based model for the phenomena of drug use, which is relatively simple, but at the same time rather realistic. 
Its concept and parameters have been established in close relation with the professional studies of the real phenomenon.
 The model permits the study of the problem at the individual and global level.
  The individual histories of the agents can be extremely diverse and appear to provide a realistic description of the real situation which could be easily compared to medical data directly.
 At the global level  we can identify the crucial most important parameters, as well as those that play a minor role. 
 This leads to a basic understanding of the complexity of the phenomenon and to the possibility of analyzing how the situation changes if the parameters are modified. This point is essential in defining the strategies for controlling and reducing the phenomenon.
 
 One of the main conclusions of our studies is the fact that drug use is usually trigged by a rare event in the personal experience.
 This shows the importance of the tail of these heterogeneous experiences, which allows us to predict that the introduction of additional heterogeneity  in the structure of the society will lead to important effects. In particular, our model's interactions among agents are assumed to be similar on average. Therefore we expect that the introduction of a complex network distribution of this iteration will be relevant. In this perspective we can already predict that the hubs of this distribution will be very important for policymaking.
 
 The model is meant to represent a realistic basis of analysis and discussion, and can be easily generalized with a complex network of social interactions and with multiple drugs with different social characteristics.
 We believe that these studies can lead to a higher level of understanding of these phenomenona and can be useful for a more effective policy making.

\section*{Methods}

\subsection*{Model parameters specification}

Following we explain in detail the parameters that characterize the agents.

\subsubsection*{Personal exposure}
The personal exposure $e_i$, describes every possible contact between the agent and the drug world and in particular describes how agents feel about drugs in general and their consumption in particular. It is defined by:
\begin{equation}
  \label{eq:pexposition}
  e_{i}(t)=\frac{\left [ {\sum\limits^{N}_{k=1}\delta{\scriptstyle (S_{k}(t)=1,2,3,4)}}\right ]-\delta{\scriptstyle (S_{i}(t)=1,2,3,4)}}{N-1}
  \end{equation}
  This expression simply counts the fraction of drug users within the entire population, but does not include the single agent under observation.
The latter represents a type of mean field interaction, which could be easily generalized to complex network situations \cite{17,15,16}.
  
\subsubsection*{Age function}

The function $\alpha(y_i(t))$ is dependent on the agent's age and its purpose is  to increase the barrier for older agents:
\begin{equation}
  \label{eq:alfay}
 \alpha(y_i(t))=\frac{\exp\left(\frac{-(-c+\lg(y_i(t)))^{2}}{2d^{2}}\right)}{ad}
 \end{equation}
The values of $a=1; d=0.5; c=3$ are chosen in order to outline the importance of age in the consumption of drugs \cite{4a}. This expression is a simple mathematical representation of how age influences the probability of drug assumption Fig. 5a. The main characteristic of this function is that the peack of this function is located between sixteen and twenty-four years and then it slowly decreases, as it has been shown by the studies of Prevo.Lab. As we can see from the Figure 5b the $\alpha(y_i(t))$ increases the average value of the $b_i^*$ due to age behavior.
 
 \begin{figure}[ht]
 \centering
\includegraphics[width=1\textwidth]{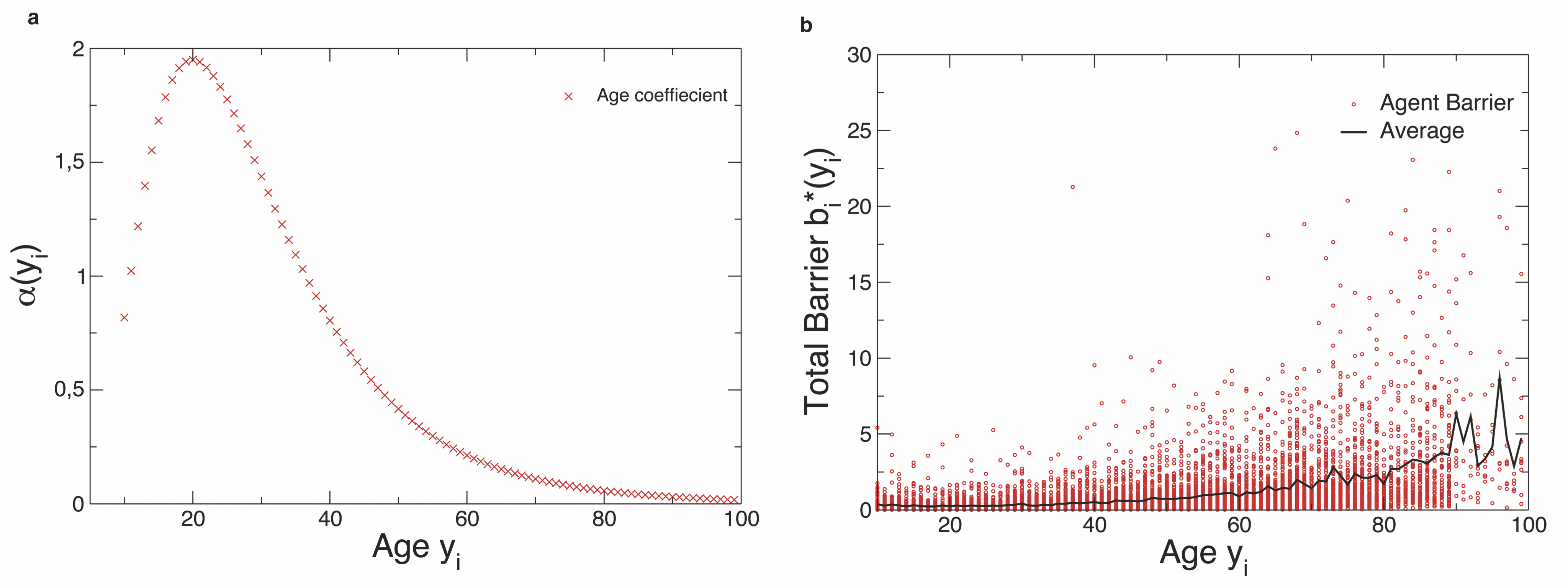}
\caption{\scriptsize\textbf{Studies of barrier age dependence. a. }The function age coefficient  $(\alpha(y_{i}))$ is described by a simple analytical form which is a reproduction of Prevo.Lab's data. \textbf{b. }This is an example of a distribution of the intrinsic agent barrier with age dependence $b_{i}^{*}(y_{i})$ for a simulation of $10000$ agents.}
\end{figure}

\subsubsection*{Drug addiction}

The drug addiction describes how an agent becomes addicted to drugs. This property is defined by the function dependence  $d_i(\pi_i(t),S_i(t))$ which has the purpose to decrease the barrier $b_i^*(t)$. The function $\pi_i(t)$ \emph{``stage permanence"} is an experience counter:
\begin{equation}
\label{eq:permanenza}
\left\{ \begin{array}{lcr}
\pi_{i}(t)=1&     \: \: \:\mbox{if} & \:   S_{i}(t)\neq S_{i}(t-1)\\
\pi_{i}(t)=\pi_{i}(t-1)+1&  \: \:\: \mbox{if} & \: S_{i}(t)=S_{i}(t-1)
\end{array} \right.
\end{equation}
and the function dependence is:
\begin{eqnarray}
\label{eq:dependence}
d_{i}(\pi_{i}(t),S_{i}(t))& = & l(t)(\exp(\pi_{i}(t)S_{i}(t)*\chi(z))-1)+d_{i}(\pi_{i}(t-1),S_{i}(t-1))
\end{eqnarray}

This expression represents an exponential growth of the addiction parameter with the persistence of the certain drug user stage.  The term $\chi(z)$ represents the \emph{drug coefficent}, which describes how addictive the drug of type $z$ is. In this regard, it would be easy to introduce drugs of different types. For the present study we consider a single case in which $\chi(z)=.0001$. This parameter could be changed for increasing or decreasing the effect of the drug addiction. The parameter  $l_i(t)$ is the \emph{first experience coefficient}, which describes the quality of the agent's first  experience in using the given drug:
\begin{equation}
\label{eq:firstexpirience}
\left\{ \begin{array}{lcc}
l_i(t)=-2&     \: \: \:\mbox{if} & \:  k_i(t)>0,75\:\&\: S_{i}(t)<3\\
l_i(t)=1&  \: \:\: & \:  \mbox{Otherwise}
\end{array} \right.
\end{equation}
where $k_i(t)$ is a random number  with uniform distribution between $[0:1]$. The two possible values of $l_i(t)$ are chosen to give a mathematical representation of the social hypothesis, that a negative experience has more psychological impact than an equivalent positive one \cite{19}. 
We can redefine the barrier of inclination to drugs with $b_{i}(t)=b_{i}^{*}(t)-d_{i}(\pi_{i}(t),S_{i}(t))$ in order to outline the role of dependence and first experience in the drug consumption and addiction.

\subsection*{Rules for the stage transitions}

We define the rules that enable agents to have stage transitions from stage $S_i(t)$ to $S_i(t+1)$. The Figure 1a in the paper shows all the possible transition stages. Each transition has some different condition due to the initial stage $S_i(t)$.

\subsubsection*{From $S_i(t)=0$}
From the stage $S_{i}(t)=0$ we can have stage transitions into:
\begin{itemize}
\item $S_i(t+1)=1$ if the EB and SEB barrier are verified for an agent with age $y_{i}(t)<26$;
\item  $S_i(t+1)=2$ if the EB and SEB barrier are verified for an agent with age $y_{i}(t)\ge26$;
\item otherwise the agent remains in the stage of non user $S_{i}(t)=0$.
\end{itemize}
As we can see from Eq.\ref{eq:s0->s1,s2} the difference between the final stage reached by the agent is dependent only on age:

\begin{equation}
\label{eq:s0->s1,s2}
\left\{ 
\begin{array}{lcc}
S_{i}(t+1)=1&\mbox{if}&[ y_{i}<26\mbox{ \& }\frac{p_{m}(t)}{m_{i}(y_{i})}<1-\gamma_{i}\mbox{ \& }b_{i}(t)+r_{i}(t)<e_{i}(t)]\\
S_{i}(t+1)=2&\mbox{if}&[y_{i}\ge26\mbox{ \& }\frac{p_{d}(t)}{m_{i}(y_{i})}<1-\gamma_{i}\mbox{ \& }b_{i}(t)+r_{i}(t)<e_{i}(t)]\\
S_{i}(t+1)=0& &\mbox{Otherwise}
\end{array}\right.
\end{equation}

\subsubsection*{From $S_i(t)=1$}
Only an agent with $y_i(t)<26$ can reach this stage. The possible transitions as we show in Eq.\ref{eq:s1->s0,s2} are:
\begin{itemize}
\item Becomes non-user, if EB or SEB are not verified;
\item Persists in the stage of mini-user if the agent gets the sufficient daily noise (opportunity) and has a budget;
\item Change in $S_i(t+1)=2$ if the agent has funds to purchase the full dose and the SEB is verified without need of the daily noise. The possibility for an agent to get in contact with the full dose is dependent on personal behavior associated with dependence level rather than the daily occasions. 
\end{itemize}

\begin{equation}
\label{eq:s1->s0,s2}
\left\{ 
\begin{array}{lcrcl}
S_{i}(t+1)=2&\mbox{ if }& [{p_{d}(t)}{m_{i}(y_{i})}<1-\gamma_{i}&\mbox{\&}&b_{i}(t)<e_{i}(t)]\\
S_{i}(t+1)=0&\mbox{ if }& [{p_{m}(t)}{m_{i}(y_{i})}>1-\gamma_{i}&\mbox{or}&b_{i}(t)+r_{i}(t)>e_{i}(t)]\\
S_{i}(t+1)=1& &\mbox{Otherwise}
\end{array}\right.
\end{equation}

An agent becomes $S_{i}(t+1)=2$ when the function dependence and the first experience coefficient decreases the SEB.

\subsubsection*{From $S_i(t)=2$}
In this stage the agent is an occasional user of normal dose, the agent can:
\begin{itemize}
\item Return to non-user stage, if EB or SEB are not verified;
\item Stay in $S_i(t+1)=2$ if both conditions are verified;
\item Become a frequent user if the agent's inclination barrier to drug is reduced by the dependence (SEB without daily noise).
\end{itemize}
The following equation shows the transition possibilities:

\begin{equation}
\label{eq:s2->s0,s3}
\left\{ 
\begin{array}{lcrcl}
S_{i}(t+1)=0&\mbox{ if }&[\frac{p_{d}(t)}{m_{i}(y_{i})}>1-\gamma_{i}&\mbox{or}&b_{i}(t)+r_{i}(t)>e_{i}(t)]\\
S_{i}(t+1)=3&\mbox{ if }&[\frac{p_{d}(t)}{m_{i}(y_{i})}<1-\gamma_{i}&\mbox{\&}&b_{i}(t)<e_{i}(t)]\\
S_{i}(t+1)=2& &\mbox{Otherwise}
\end{array}\right.
\end{equation} 
As we can see, the transition rules are quite similar to the previous stage. We have considered that the stage $S_{i}(t)=2$ is reachable by all agents without being contingent on age.

\subsubsection*{From $S_i(t)=3$}
The agent is a frequent user. She does not care about the EB because the addiction raises the necessity of a dose and the money becomes a negligible constraint. She can become:
\begin{itemize}
\item Heavy user, pathological, if her inclination barrier to drug decreases by a factor $5$ or more;
\item She can decide to stop using drugs if the social environment changes drastically, or with the $10\%$ probability, which represents the possibility to stop using drugs due to exogenous environmental factors.
\item Otherwise the agent $i$ continues to take drug as a frequent user.
\end{itemize}   
Here we show the transition rules:
\begin{equation}
\label{eq:s3->s0,s4}
\left\{ 
\begin{array}{lcc}
S_{i}(t+1)=4&\mbox{ if }&5b_{i}(t)<e_{i}(t) \\
S_{i}(t+1)=0&\mbox{ if }&[b_{i}(t)>e_{i}(t)\mbox{ or } c_{i}(t)>0.9]\\
S_{i}(t+1)=3&& \mbox{Otherwise}
\end{array}\right.
\end{equation}
Where $c_i(t)$ is a random number from uniform distribution $[0:1]$.
There is no more the daily noise $r_{i}(t)$ and the transitions depend on the variation of $b_{i}(t)$ due to the dependence function and the stage persistence. The factor $5$ of the first condition is based on our decision \cite{4,4a,4c} to discriminate in a significant way the difference between the two stages $S_{i}(t)=3$ and $S_{i}(t+1)=4$.
 The $b_{i}(t)>e_{i}(t)$ therefore represents the decreasing possibility of the exposure factor due to a sudden change in the social interactions.

\subsubsection*{From $S_i(t)=4$}

When an agent becomes a heavy user, she has only two possible choices according to the transition table showed in the main article. The possibilities are:
\begin{itemize}
\item The agent can leave drug with the probability of $5\%$;
\item The agent can persist in the stage $S_i(t+1)=4$.
\end{itemize}
At this stage, the situation of the agent is independent of the condition  EB and SEB due to every addiction. Therefore we assume only a simple probabilistic transition:
\begin{equation}
\label{eq:s4->s0,s4}
\left\{ 
\begin{array}{lcc}
S_{i}(t+1)=0&\mbox{if}&l_{i}(t)>0.95\\
S_{i}(t+1)=4&& \mbox{Otherwise}
\end{array}\right.
\end{equation}
Where $l_i(t)$ is a random number from uniform distribution $[0:1]$.
The probability of the $5\%$ is an estimation suggested by the Prevo.lab studies and is related to the possibility of recovering from drug addiction.

In this way we have defined all the rules that allow an agent to evolve and reach all possible stages of consumption.
So far, the agent cannot die and the age $y_{i}(t)$ is extracted at the beginning of the simulation and remains constant for every $t$. Later we define the rules that describe the death and birth of the agent and its aging dynamics.

\subsection*{Aging}

Here we show the rules that define the process of an agent's vital cycle (death and birth), and the dynamic that leads an agent to overdose.

\subsubsection*{Death and birth condition}

Based on the 2007 ISTAT data we consider the probability of death in a given year for a person of a given age. We label this probability $P_{y_i(t)}$ and every $\tau=50t$ consecutive iteration we check this rule for each of $N$ agents: 
\begin{equation}
\label{eq:death}
\left\{ 
\begin{array}{lr}
u_i(t)<P_{y_i(t)}&y_i(t+1)=y_i(t)+1\\
\mbox{Otherwise}&\mbox{$i$ dies }y_i(t+1)=10\\
\end{array}\right.
\end{equation}
Where $u_i(t)$ is a random number from uniform distribution $[0 : 1]$. 
When the agent $i$ dies, she will be replaced by another agent with age $y_i(t+1)=10$ and she will be characterized by the new extraction of the parameters  $\beta_i, \gamma_i$ from the current value of the distributions at time $t$. Instead, if she lives, she will be one year older.

\subsubsection*{Overdose rule}

By adding the possibility of death, it is possible to improve also some rules of the $S_i(t)=4$ stage transition. An agent in $S_{i}(t)=4$ can become a non-user by either detoxifying in a treatment center or by dying of overdose. When the agent dies of an overdose, the agent will be replaced by a new agent with the new extraction of the parameters  $\beta_i, \gamma_i$ from the current value of the distributions at time $t$. The following equation expresses the overdose death condition:
\begin{equation}
\label{eq:s4new}
\left\{ 
\begin{array}{lcl}
S_{i}(t+1)\rightarrow0&\mbox{ if }g_t(i)>0.1&\mbox{lives}\\
S_{i}(t+1)\rightarrow0&\mbox{Otherwise}&\mbox{dies}\\
\end{array}\right.
\end{equation}
Where $g_i(t)$ is a random number from uniform distribution $[0 : 1]$. The agent with the probability of $10\%$ dies. Otherwise she lives. The probability to survive after an overdose is taken from the data of SimDrug and  Sert \cite{4c,10}. With our dynamics we can replicate the same overdoses trend of SimDrug model too.

\section*{Acknowledgements}

We acknowledge the study group of Prevo.lab Project  in particular Riccardo Gatti, for many interesting and important discussions.
This paper is partly supported by the PNR Project CRISIS-lab.

\section*{Author contributions}
R.D.C. \& L.P. designed and developed the model. R.D.C. made the computer simulations. All authors contributed evenly to the manuscript.

\end{document}